# A Survey of Data Mining Techniques for Social Network Analysis

Mariam Adedoyin-Olowe[1], Mohamed Medhat Gaber[1] and Frederic Stahl[2]

[1]School of Computing Science and Digital Media, Robert Gordon University
Aberdeen, AB10 7QB, UK
[2]School of Systems Engineering, University of Reading
PO Box 225, Whiteknights, Reading, RG6 6AY, UK

**Abstract.** Social network has gained remarkable attention in the last decade. Accessing social network sites such as Twitter, Facebook LinkedIn and Google+ through the internet and the web 2.0 technologies has become more affordable. People are becoming more interested in and relying on social network for information, news and opinion of other users on diverse subject matters. The heavy reliance on social network sites causes them to generate massive data characterised by three computational issues namely; size, noise and dynamism. These issues often make social network data very complex to analyse manually, resulting in the pertinent use of computational means of analysing them. Data mining provides a wide range of techniques for detecting useful knowledge from massive datasets like trends, patterns and rules [44]. Data mining techniques are used for information retrieval, statistical modelling and machine learning. These techniques employ *data pre-processing, data analysis, and data interpretation* processes in the course of data analysis. This survey discusses different data mining techniques used in mining diverse aspects of the social network over decades going from the historical techniques to the up-to-date models, including our novel technique named **TRCM**. All the techniques covered in this survey are listed in the Table.1 including the tools employed as well as names of their authors.

## 1. Introduction

Social network is a term used to describe web-based services that allow individuals to create a public/semi-public profile within a domain such that they can communicatively connect with other users within the network [22]. Social network has improved on the concept and technology of Web 2.0, by enabling the formation and exchange of User-Generated Content [46]. Simply put, social network is a graph consisting of *nodes* and *links* used to represent social relations on social network sites [17]. The *nodes* include entities and the relationships between them forms the *links* (as presented in Fig. 1).

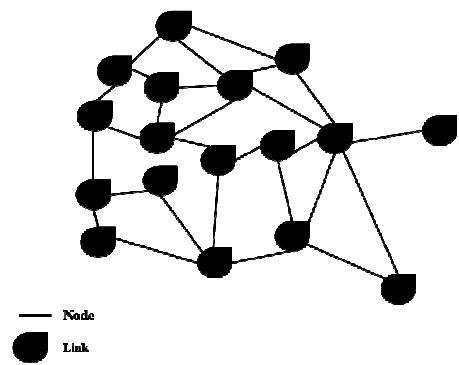

**Fig. 1. Social Network showing nodes and links**

Social networks are important sources of online interactions and contents sharing [81], [21], subjectivity [6], assessments [52], approaches [54], evaluation [48], influences [8], observations [24], feelings [46], opinions and sentiments expressions [66] borne out in text, reviews, blogs, discussions, news, remarks, reactions, or some other documents [57]. Before the advent of social network, the homepages was popularly used in the late 1990s which made it possible for average internet users to share information. However, the activities on social network in recent times seem to have transformed the *World Wide Web (www)* into its intended original creation. Social network platforms enable rapid information exchange between users regardless of the location. Many organisations, individuals and even government of countries now follow the activities on social network. The network enables big organisations, celebrities, government official and government bodies to obtain knowledge on how their audience reacts to postings that concerns them out of the enormous data generated on social network (as shown in Fig. 2). The network permits the effective collection of large-scale data which gives rise to major computational challenges. However, the application of efficient data mining techniques has made it possible for users to discover valuable, accurate and useful knowledge from social network data.

Data mining techniques have been found to be capable of handling the three dominant disputes with social network data namely; **size, noise and dynamism**. The voluminous nature of social network datasets require automated information processing for analysing it within a reasonable time. Interestingly, data mining techniques also require huge data sets to mine remarkable patterns from data; social network sites appear to be perfect sites to mine with data mining tools [27]. This forms an enabling factor for advanced search results in search engines and also helps in better understanding of social data for research and organizational functions [4]. Data mining tools surveyed in this paper ranges from unsupervised, semi-supervised to supervised learning. A table itemizing the techniques covered in this paper is presented in section 7.

The rest of the survey is organised as follows. Section 2 examines the social network background. Section 3 listed research issues on social network analysis. Section 4 discusses some of the Graph Theoretic tools used for social network analysis. Section 5 gives an overview of tools used to analyse opinions conveyed on social network while Section 6 presents some of the sentiment analysis techniques used on social network. Section 7 describes some unsupervised classification techniques employed in social network analysis**.** Section 8 presents topic detection and tracking tools used for social network analysis. The survey is concluded in Section 9 by stating the direction for future work.

## 2. Social Network Background

During the last decade social network have become not only popular but also affordable and universally-acclaimed communication means that has thrived in making the world a global village. Social network sites are commonly known for information dissemination, personal activities posting, product reviews, online pictures sharing, professional profiling, advertisements and opinion/sentiment expression. News alerts, breaking news, political debates and government policy are also posted and analysed on social network sites. It is observed that more people are becoming interested in and relying on the social network for information in real time. Users sometimes make decisions based on information posted by unfamiliar individuals on social network [66] increasing the degree of reliance on the credibility of these sites. Social network has succeeded in transforming the way different entities source and retrieve valuable information irrespective of their location. Social network has also given users the privilege to give opinions with very little or no restriction.

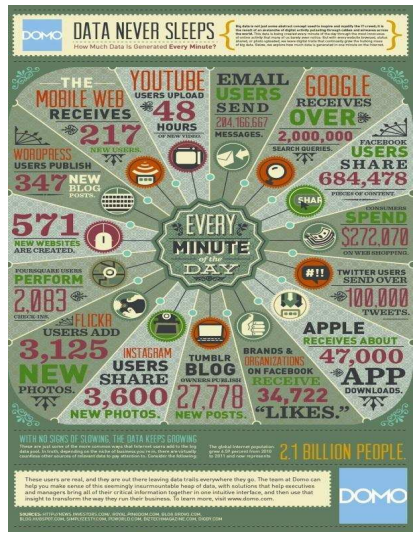

**Fig.2. Estimated Data Generated on Social Network Sites Every Minute**
*Thumbnail image courtesy of iStockphoto, loops7*
http://mashable.com/2012/06/22/data-created-every-minute/

### 2.1 Social Network – Power to the Users

Social sites have undoubtedly bestowed unimaginable privilege on their users to access readily available never-ending uncensored information. Twitter, for example, permits its users to post events in real time way ahead the broadcast of such events on traditional news media. Also, social network allow users to express their views, be it positive or negative [4]. Organizations are now conscious of the significance of consumers' opinions posted on social network sites to the patronage of their products or services and the overall success of their organisations. On the other hand, important personalities such as celebrities and government officials are being conscious of how they are perceived on social network. These entities follow the activities on social network to keep abreast with how their audience reacts to issues that concerns them [20], [23], [40]. Considering the enormous volume of data being generated on social network, it is imperative to find a computational means of filtering, categorising, classifying and analysing the social network contents.

### 3. Research Issues on Social Network Analysis

A number of research issues and challenges facing the realisation of utilising data mining techniques in social network analysis could be identified as follows:

- **Linkage-based and Structural Analysis –** This is an analysis of the linkage behaviour of the social network so as to ascertain relevant nodes, links, communities and imminent areas of the network - Aggarwal, 2011.

- **Dynamic Analysis and Static Analysis** – Static analysis such as in bibliographic networks is presumed to be easier to carry out than those in streaming networks. In static analysis, it is presumed that social network changes gradually over time and analysis on the entire network can be done in batch mode. Conversely, dynamic analysis of streaming networks like Facebook and YouTube are very difficult to carry out. Data on these networks are generated at high speed and capacity. Dynamic analysis of these networks are often in the area of interactions between entities - Papadopoulos et al, (2012), temporal events on social networks Adedoyin-Olowe et al (2013); Becker et al (2011) and evolving communities - Fortunato, (2010).

Having presented some of the research issues and challenges in social network analysis, the following sections and sub-sections present the overview of different data mining approaches used in analysing social network data.

## 4 Graph Theoretic

Graph theory is probably the main method in social network analysis in the early history of the social network concept. The approach is applied to social network analysis in order to determine important features of the network such as the *nodes* and *links* (for example *influencers* and *the followers*). Influencers on social network have been identified as users that have impact on the activities or opinion of other users by way of followership or influence on decision made by other users on the network (as presented in Fig.3). Graph theory has proved to be very effective on large-scale datasets (such as social network data). This is because it is capable of bye-passing the building of an actual visual representation of the data to run directly on data matrices [76]. In [19] *centrality measure* was used to inspect the representation of power and influence that forms clusters and cohesiveness [16] on social network. The authors of [34] employed *parameterized centrality metric* approach to study the network structure and to rank nodes connectivity. Their work formed an extension of *a-centrality* approach which measures the number of alleviated paths that exist among nodes.

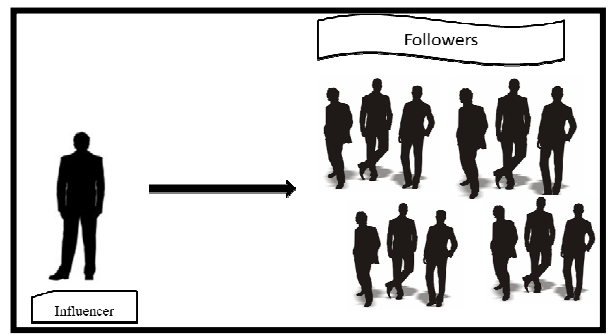

**Fig 3. Graph Theoretic on Social Network**

### 4.1 Community Detection Using Hierarchical Clustering

A community is a smaller compressed group within a larger network (as shown in Fig.4). Community formation is known to be one of the important characteristics of social network sites. Users with similar interest form communities on social network thereby displaying strong sectional structure. Communities on social networks, like any other communities in the real world, are very complex in nature and difficult to detect. Applying the appropriate tools in detecting and understanding the behaviour of network communities is crucial as this can be used to model the dynamism of the domain they belong [4]. Different authors have applied diverse clustering techniques to detect communities on social network [35]; [32]; 68], with *hierarchical clustering* being mostly used [62]. This technique is a combination of many techniques used to group nodes in the network to reveal strength of individual groups which is then used to distribute the network into communities. **Vertex clustering** belongs to hierarchical clustering methods, graph vertices can be resolved by adding it in a vector space so that pairwise length between vertices can be measured. ***Structural equivalence measures*** of hierarchical clustering concentrate on number of common network connections shared by two nodes. Two people on social network with several mutual friends are more likely to be closer than two people with fewer mutual friends on the network. Users in the same social network community often recommend items and services to one another based on the experience on the items or services involved. This is known as ***recommender system*** as explains next in section 4.2.

### 4.2   Recommender System in Social Network Community

Based on the mutuality between nodes in social network groups, ***collaborative filtering (CF)*** technique, which forms one of the three classes of the ***recommender system (RS)***, can be used to exploit the association among users [56]. Items can be recommended to a user based on the rating of his mutual connection. Where *CF's* main downside is that of data sparsity, ***content-based*** (another *RS* method) explore the structures of the data to produce recommendations. However, the *hybrid approaches* usually suggest recommendations by combining *CF* and *content-based* recommendations. The experiment in [18] proposed a hybrid approach named ***EntreeC***, a system that pools *knowledge-based RS* and *CF* to recommend restaurants. The work in [69] improved on *CF* algorithm by using a greedy implementation of ***hierarchical agglomerative clustering*** to suggest forthcoming conferences or journals in which researchers (especially in computer science) can submit their work.

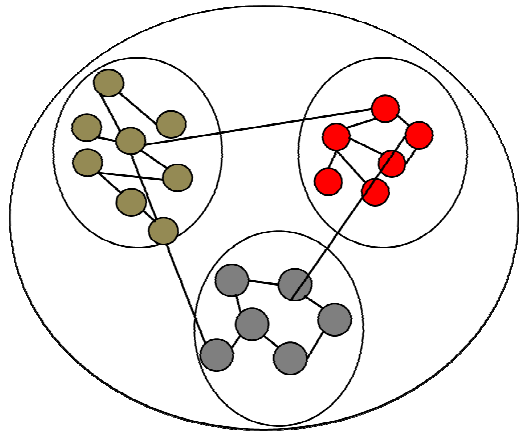

**Fig. 4 Social Network Community Structure**

### 4.3   Semantic Web of Social Network

The ***Semantic Web*** platform makes knowledge sharing and re-use possible over different applications and community edges. Discovering the evolvement of ***Semantic Web*** (***SW***) enhances the knowledge of the prominence of ***Semantic Web Community*** and envisages the synthesis of the Semantic Web. The work in [92] employed ***Friend of a Friend (FOAF)*** to explore how local and global community level groups develop and evolve in large-scale social networks on the *Semantic Web*. The study revealed the evolution outlines of social structures and forecasts future drift. Likewise [74] application model of ***Semantic Web-based Social Network Analysis Model*** creates the ontological field library of social network analysis combined with the conventional outline of the semantic web to attain intelligent retrieval of the Web services. Furthermore, *VoyeurServer* [61] improved on the open-source *Web-Harvest framework* for the collection of online social network data in order to study structures of trust enhancement and of online scientific association. *Semantic Web* is a relatively new area in social network analysis and research in the field is still evolving.

## 5        Opinion Analysis on Social Network

According to Technorati, about 75,000 new blogs and 1.2 million new posts giving opinion on products and services are generated every day [50]. Also massive data generated every minute on common social network sites (as presented in Fig. 2) are laden with opinion of users as regards diverse subject ranging from personal to global issues [80]. Users' opinions on social network sites can be referred to as discovery and recognition of positive or negative expression on diverse subject matters of interest. These opinions are often convincing and their indicators can be used as motivation when making choices and decisions on patronage of certain products and services or even endorsement of political candidate during elections [47], [67]. Even though online opinions can be discovered using traditional methods, this form is conversely inadequate considering the large volume of information generated on social network sites. This fact underscores the relevance of data mining techniques in mining opinion expressed on social network site.

Various methods have been developed to analyse the opinion arising from products, services, events or personality review on social network [36]. Data mining tools already used for opinion and sentiment analysis include collections of simple counting methods to machine learning. Categorizing opinion-based text using binary distinction of positive against negative [39], [29], [64], [85], is found to be insufficient when ranking items in terms of recommendation or comparison of several reviewers' opinions [65] (e.g., using actors starred in two different films to decide which of them to see at the cinema). Determining players from documents on social network has also become valuable as influential actors are considered as variables in the documents [91] when applying data mining techniques on social network. The idea of co-occurrence can also be seen as viable information [34].

Data mining techniques used for opinion mining on social network are discussed in the next section of this survey.

### 5.1     Aspect-Based/Feature-Based Opinion Mining

*Aspect-based* also known as *feature-based analysis* is the process of mining the area of entity customers has reviewed [41]. This is because not all aspects/features of an entity are often reviewed by customers. It is then necessary to summarise the aspects reviewed to determine the polarity of the overall review whether they are positive or negative. Sentiments expressed on some entities are easier to analyse than others, one of the reason being that some reviews are ambiguous.

According to [57] aspect-based opinion problem lies more in blogs and forum discussions than in product or service reviews. The aspect/entity (which may be a computer device) reviewed is either *'thumb up'* or *'thumb down'*, thumb up being positive review while thumb down means negative review. Conversely, in blogs and forum discussions both aspects and entity are not recognized and there are high levels of insignificant data which constitute noise. It is therefore necessary to identify opinion sentences in each review to determine if indeed each opinion sentence is positive or negative [41]. Opinion sentences can be used to summarize aspect-based opinion which enhances the overall mining of product or service review.

An opinion holder expresses either positive or negative opinion [13], [51], [89] on an entity or a portion of it when giving a regular opinion and nothing else [41]. However, [53] put necessity on differentiating the two assignments of finding out neutral from non-neutral sentiment, and also positive and negative sentiment. This is believed to greatly increase the correctness of computerised structures.

### 5.2   *Homophily* Clustering in Opinion Formation

Opinion of influencers on social network is based largely on their personal views and cannot be hold as absolute fact. However, their opinions are capable of affecting the decisions of other users on diverse subject matters. Opinions of influential users on Social network often count, resulting in opinion formation evolvement. Clustering technique of data mining can be utilised to model opinion formation by way of assessing the affected nodes and unaffected nodes. Users that depict the same opinion are linked under the same nodes and those with opposing opinion are linked in other nodes (as shown in Fig.5). This concept is referred to as *homophily* in social network [59]. Homophily can also be demonstrated using other criteria such as race and gender [42].

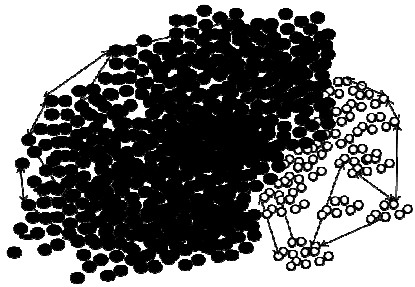

**Fig. 5.  Homophily Showing Influencer and Followers' Opinion in Social Network**

Behaviour of participants in each node is subject to adjustments in awareness of the behaviour of participants in other nodes [47]. Opinion formation starts from the *initial* stage where bulk of participants pays no attention to recourse action on significant issue at this stage. This is so because they do not consider the action viable. When cogent information is introduced opinion is cascaded and participants begin to make either positive or negative decisions. At this stage the decision of influential participants who are either efficient in the field or in communication skill attracts the followership of the minority. The initial stage (as presented Fig. 6) is then transformed to *alert* stage (in Fig. 7).  The percolate stage sets in when the minority are able to form a different opinion based on other agents' behaviour and introduction of new information. It is worthy of note that opinion of social network users is oftentimes mix with sentiment; sentiment analysis on social network is discussed next in section.6.

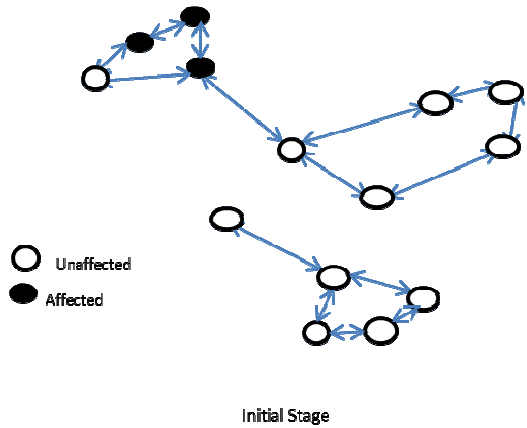

**Fig.6. Initial stage of Opinion Formation**

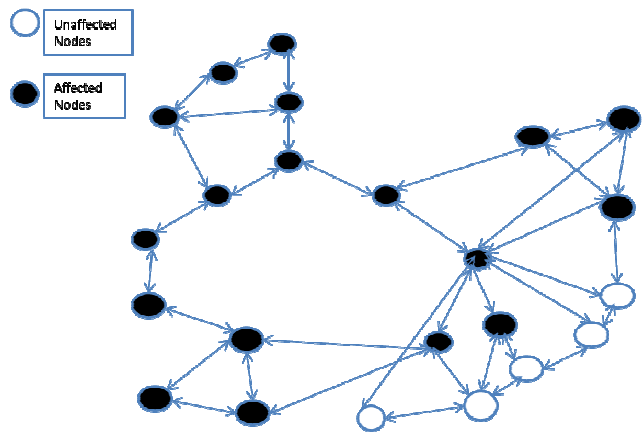

**Fig. 7 Alert State of Opinion Formation**

### 5.3  Opinion Definition and Opinion Summarization

Extensive information gives rise to challenge of automatic summarization. ***Opinion definition*** and ***opinion summarization*** are essential techniques for recognizing opinion. Opinion definition can be located in a text, sentence or topic in a document; it can also reside in the entire document. Opinion summarization sums up different opinions aired on piece of writing by analysing the sentiment polarities, degree and the associated occurrences. Authors in [55] used ***Support Vector Machine (SVM)*** with ***linear kernel*** to learn the polarity of neutral examples in documents. Their results propose that polarity complications can be adequately handled as three class complications using pairwise coupling while merging outcomes in remarkable ways. ***Opinion extraction*** is vital for summarization and subsequent tracking. Texts, topics and documents are searched to extract opinionated part. It is necessary of summarise opinion because not all opinion expressed in a document are expected to be of importance to issue under consideration. Opinion summarization is quiet useful to businesses and government alike, as it helps in improving policies and products respectively [29], [60]. It also helps in taking care of information excess [87].

### 5.4 Opinion Extraction

Sentiment analysis deals with establishment and classification of subjective information present in a material [86]. This might not necessarily be fact-based as people have different feelings toward the same product, service, topic, event or person. Opinion extraction is necessary in order to target the exact part of the document where the real opinion is expressed. Opinion from an individual in a specialised subject may not count except if the individual is an authority in the field of the subject matter. Nevertheless, opinion from several entities necessitates both opinion extraction and summarization [57].

In opinion extraction, the more the number of people that give their opinion on a particular subject, the more important that portion might be worth extracting. Opinion can aim at a particular article while on the other hand can compare two or more articles. The formal is a regular opinion while the latter is comparative [43]. Opinion extraction identifies subjective sentences with sentimental classification of either positive or negative.

## 6 Sentiment Analysis on Social Network

Sentiment analysis research has its roots in papers published by [28] and [83] where they analysed market sentiment. The concept later gained more ground the year after where authors like [64] and [85] have reported their findings. Sentiment analysis can be referred to as discovery and recognition of positive or negative expression of opinion by people on diverse subject matters of interest. Opinions expressed by social network users are often convincing and these indicators can be used to form the basis of choices and decisions made by people on patronage of certain products and services or endorsement of political candidate during elections [47], [67].

It is worthy of note that the enormous opinions of several millions of social network users are overwhelming, ranging from very important ones to mere assertions (e.g. *"The phone does not come in my favourite colour, therefore it is a waste of money"*). Consequentially it has become necessary to analyse sentiment expressed on social network with data mining techniques in order to generate a meaningful frameworks that can be used as decision support tools. Diverse algorithms are employed to ascertain sentiment that matters to a topic, text, document or personality under review. The purpose of sentiment analysis on social network is to recognize potential drift in the society as it concerns the attitudes, observations, and the expectations of stakeholder or the populace. This recognition enables the entities concern to take prompt actions by making necessary decisions. It is important to translate sentiment expressed to useful knowledge by way of mining and analysis.

Having given an overview of sentiment analysis on social network, an overview of some of the data mining tools used for sentiment analysis on social network are discussed in subsequent sections of the survey.

### 6.1 Sentiment Orientation (SO)

Widespread products are likely to attract thousands of reviews and this may make it difficult for prospective buyers to track usable reviews that may assist in making decision. On the other hand sellers make use of **Sentiment Orientation (SO)** for their rating standard in other to safeguard irrelevant or misleading reviews present to reviewers the *5-star scale rating* with *five* signifying best rated while *one* signifies poor rating. In [49] **SO** was used to improve the performance of mood classification. Livejournal blog corpus dataset was used to train and evaluate the method used. The experiment presented a modular proficient ***hierarchical classification*** technique easily implemented together with *SO* attributes and machine learning techniques. The initial result of classification accuracy however recorded slightly above the baseline. An incorporation of ***flexible hierarchy-based mood*** approach to mood classification discovers that attributes that points to accurate classification of mood expression can be retrieved from the various dense blog corpus domain.

### 6.2  Product Ratings and Reviews

The reliance on the internet (especially social network sites) for information when making choices about products or services has increased the necessity of researching into the electronic-word-of-mouth. Products (services) ratings and reviewing often contains sentiment expressions [33]; [79], an item can be rated based on the mood of the reviewer at the time [15]. Social network sites such as *Epinions and Ciao* allow users to establish a trust network among them showing who to trust in offering product reviews and ratings. Most online stores equally afford their customers the opportunity to either rate/review product or service they purchased. This process enables prospective customers to have access to first-hand information about these products/services before making purchase. Poorly rated or reviewed products/services tend to attract very low patronage or no patronage at all.

  Data mining tools are used to analyse the concept of products ratings and reviews on social network. The experiments in [7] proposed an advanced ***matrix factorisation method*** capable of increasing rating predictions and estimate accurate strengths of trust associations within the same period. Their work suggests that even though general users who trust others in the same network tend to have similar ratings over time (*homophily and social influence*), this does not implies similar preferences. Furthermore, [78] proposed a simplified context that exploits multi-modal social networks that provide item recommendations in *social rating networks* (*SRNs*). Using their ***Social-Union*** method which associates similarity matrices originated from heterogeneous (unipartite and bipartite) unequivocal or implied *SRNs,* they concluded that *social-union* out-performed existing renowned item recommendation systems.

### 6.3  Reviews and Ratings (RnR) Architecture (Rahayu, 2010)

  ***RnR*** is a conceptual architecture created as an interactive structure. It is user input oriented that develops relative new reviews. The user supplies the name of product or service whose performance has been earlier reviewed online by customers. This systems checks through the corpus of already reviewed to find out if it is stored in the local cache of the architecture for latest reviews. If the supplied data is found to be recent, it is subsequently used. If it is found to be obsolete, crawling is done on secondary site such as TripAdvisor.com and Expedia.com.au. The data retrieved on these sites are then locally built to discover the required justification. ***RnR*** architecture produced complete remarks of product and service (under review) within a time line by utilising ***temporal dimension analysis*** with ***scatter plot*** and ***linear regression***. Tagging of few words was used for free accessible domain ontology for feature identification. This is because tagging ***POS*** of each word in whole reviews and opinion word identification process could be time consuming and computationally expensive even though it produces high accuracy. The use of *neighbour word* (words around the feature occurrences) also aids the pruning down of computational overhead.

### 6.4 Aspect Rating Analysis

Aspect-rating is numerical evaluations in relation to the aspect pointing to the level of satisfaction portrayed in the comments gathered toward this aspect and the aspect rating. It makes use of diminutive phrases and their modifiers [57], for example *'good product, excellent price'*. Each aspect is extracted and collated using Probabilistic latent semantic analysis (pLSA). It can be used in place of structure of the phrase. The already known complete post is exploited to ascertain the aspect rating. Aspect cluster are words that communally stands for an aspect that users are concerned in and would comment on.

Latent Aspect Rating Analysis *(LARA)* approach attempts to analyse opinion borne by different reviewers by doing a *text mining* at the point of topical aspect. This enables the determinance of every reviewer's *latent score* on each aspects and the relevant influence on them when arriving at an affirmative conclusion. The revelation of the latent scores on different aspects can instantly sustain aspect-base opinion summarization. The aspect influences are proportional to analysing score performance of reviewers. The fusion latent scores and aspect influences is capable of sustaining personalised aspect-level scoring of entities using just those reviews originated from reviewers with comparable aspect influences to those considered by a particular user [87]. An aspect-based summarization make use of set of user reviews of a subject as input and creates a set of important aspects taking into consideration the combined sentiment of each aspect and supporting textual indication[82].

Sentiment analysis tools used for social network analysis commonly utilize sentimental words often compiled into sentiment lexicon (also know as sentiment dictionary). Sentiment lexicon is briefly discussed next in section 6.5.

### 6.5 Sentiment Lexicon

***Sentiment Lexicon*** is a dictionary of sentimental words reviewers often used in their expression. Sentiment lexicon is list of the common words that enhances data mining techniques when used mining sentiment in document. Different corpus of sentiment lexicon can be created for variety of subject matters. For instance sentimental words used in sport are often different from those used in politics. Expanding the occurrence of sentiment lexicon helps to focus more on analysing topic-specific occurrence, but with the use of high manpower [36]. Lexicon-based approaches require parsing to work on simple, comparative, compound, conditional sentences and questions [30].

Sentiment lexicon can be expanded by use of synonyms. Nonetheless, lexicon expansion through the use of synonyms has a drawback of the wording loosing it primary meaning after a few recapitulation. Sentiment lexicon can also be enhanced by 'throwing away' neutral words that depicts neither positive nor negative expression. Neutral expression is common especially in products ratings and reviews.

### 7 Unsupervised Classification of Social Network Data

A straightforward unsupervised learning algorithm can be used to rate a review as *'thumbs up'* or *'thumbs down'* [85]. This can be by way of digging out phrases that include adjective or adverbs (part of speech tagging) [75]. The semantic orientation of every phrase can be approximated using ***PMI-IR*** [120] and then classify the review using the ***average semantic orientation*** of the phrase. Cogency of title, body and comments generated from blog post has also been used in clustering similar blogs into significant groups. In this case keywords played very important role which may be multifaceted and bare [3]. ***EM-based*** and ***constrained-LDA*** were utilized to cluster aspect phrases into aspect categories [90]. In [12] two unsupervised frameworks based on link structure of the Web pages, and Agglomerative/Conglomerative Double Clustering (A/CDC) was used to find group of individuals on the web. The result proves to be more accurate than those obtained by traditional agglomerative clustering by more than 20% while achieving over 80% F-measure.

Other unsupervised learning used in sentiment analysis in products rating and reviews include **POS (Part of Speech) tagging**. In POS adjectives are tagged to display positive and negatives ones. Sentiment polarity is the binary classification of an opinionated document into a largely positive and negative opinion [67]. In review this is commonly termed with the '*thumps up*' and '*thumps down*' expressions as mentions earlier. The polarity of positive against negative is weighed to give an overall analysis of sentiment expressed on issue under review. **Bootstrapping** also forms part of the unsupervised approaches. It utilizes obtainable primary classifier to make labelled data which a supervised process can build upon [45], [73], [89]. **Semantic orientation** is also an unsupervised approaches currently used for sentiment analysis on social network. It attaches different meaning to a single word – synonym. This could either be positive or negative (for example '*the party is bad*' may in actual fact mean the party is fun). Direction and intensity of words used can determine the semantic orientation of the opinion expressed [14].Semi-supervised and supervised classification are more structured techniques as discussed next in Sections 7.1 and 7.2.

### 7.1 Semi-supervised Classification

Semi-supervised learning is a goal-targeted activity but unlike unsupervised; it can be specifically evaluated. Authors of [31] worked on a mini training set of seed in positive and negative expressions selected for training a term classifier. Synonym and antonym comparatives were added to the seed sets in an online dictionary. The approach was meant to produce the extended sets *P'* and *N'* that makes up the training sets. Other learners were employed and a binary classifier was built using every glosses in the dictionary for both term in *P' ∪ N'* and translating them to a vector. Their approach discovers the origin of information which they reported was missing in earlier techniques used for the task. **Semi-supervised lexical classification** proposed by [77] integrated lexical knowledge into supervised learning and spread the approach to comprise unlabelled data. Cluster assumption was engaged by grouping together two documents with the same cluster basically supporting the positive - negative sentiment words as sentiment documents. It was noted that the sentiment polarity of document decides the polarity of word and vice versa.

In [72] semi-supervised learning uses **polarity detection** as semi-supervised label propagation problem in graphs. Each node representing words whose polarity is to be discovered. The results shows label propagation progresses outstandingly above the baseline and other semi-supervised techniques like **Mincuts** and **Randomized Mincuts**. The work of [38] compared **graph-based semi-supervised learning** with regression and [65] proposed **metric labelling** which runs *SVM* regression as the original label preference function comparable to similarity measure. Their result shows that the graph-based semi-supervised learning (*SSL*) algorithm as per PSP (*positive-sentence-percentage*) comparison (SSL+PSP) proved to perform well.

## 7.2    Supervised Classification

While clustering techniques are used where basis of data is established but data pattern is unknown [4], classification techniques are supervised learning techniques used where the data organisation is already identified. It is worthy of mention that understanding the problem to be solved and opting for the right data mining tool is very essential when using data mining techniques to solve social network issues. Pre-processing and considering privacy rights of individual (as mentioned under research issues of this paper) should also be taken into account. Nonetheless, since social media is a dynamic platform, impact of time can only be rational in the issue of topic recognition, but not substantial in the case of network enlargement, group behaviour/ influence or marketing. This is because this attributes are bound to change from time to time. Information updates in some Social network such as twitters and Facebook present *Application Programmers Interfaces* (*APIs*) that makes it possible for crawler, which gather new information in the site, to store the information for later usage and update.

In [85] a supervised learning algorithm used the combination of *multiple bases of facts* to label couple of adjectives having similar or dissimilar semantic orientations. The algorithm resulted in a graph with nodes and links which represents adjectives and similarity (or dissimilarity) of semantic orientation respectively.

## 8    Topic Detection and Tracking on Social Network

*Topic Detection and Tracking (TDT)* on social network employs different techniques for discovering the emergent of new topics (or events) and for tracking their subsequent evolvements over a period of time. *TDT* is receiving high level of attention recently. Many researchers and authors are conducting experiments on *TDT* on social network sites, especially on Twitter [1]; [37]; [9]; [10]; [63]; [70]; [58]. In [25] *support vector machine (SVM)* was found to be efficient in training Twitter hashtags metadata when predicting the political alignment of twitter users. Authors of [9] used an *incremental online clustering* algorithm to cluster a stream of Twitter messages in real time. They trained a *Naïve Bayes-Text classifier* to distinguish between *fastest-growing real-world events* contents and *non-events* contents on Twitter. The performance of the training set shows the precision of all classifier computed in *10-fold cross-validation.* The experiments in [11] used a range of query-building approaches to automatically enhance user-contributed information for planned events with robustly generated Twitter contents. Their approach used *browser plug-in script* and a *customizable web interface* to identify relevant Twitter content for planned events.

The experiments in [5] proposed a combination of six techniques namely; **LDA (*Latent Dirichlet Allocation*)**, ***Doc-p* (*Document-Pivot Topic Detection, GFeat-p (Graph-based Feature-Pivot Topic Detection), FPM (Frequent Pattern Mining), SFPM (Soft Frequent Pattern Mining)*** and ***BNgram*** for real-world event detection on Twitter network. The techniques were verified on tweets relating to three major events (*English FA Cup Finals, US Super Tuesday Primaries* and *US Elections 2012*) with variations in time scale and topic mix level. The algorithms revealed that dataset pre-processing and sampling process affects the quality of topic retrieved. Conversely, the algorithms performed optimally on the three datasets considered. Similarly, [70] proposed an algorithm for detecting and tracking breaking news in Twitter. The application named "***Hotstream***" was built to afford its users the opportunity of detecting and tracking breaking news from Twitter timeline. Authors of [63 proposed a state-of-the-art ***First Story Detection (FSD)*** technique to detect *predictable* and *unpredictable* events using real-time indication from Wikipedia and Twitter data streams. The result of the experiments recorded about 2-hour delay for Wikipedia in real-world events. Authors in [88] used *EDCoW* (*Event Detection with Clustering of Wavelet-based Signals*) to cluster words to form events with a modularity-based graph partitioning method. On the other hand [26] employed lightweight event detection using *wavelet signal analysis* of hashtags occurrences in Twitter public stream. The experiments used *Latent Dirichlet Allocation topic inference model* based on Gibbs Sampling. The outcome of the experiments shows that peak detection using *Continuous Wavelet Transformation* realized impressive outcomes in the ascertaining abrupt increases on the mention of specific hashtags. In [1] the abruptness in hashtags usage is labelled *unexpected rule evolvement* which is discussed under *TRCM* in Section 8.1.

## 8.1 *TRCM* for TDT

Twitter as a social network and hashtags as tweet labels can be analysed in order to detect changes in event patterns (TDT) using Association Rules (*ARs*). Twitter data can be used to analyse patterns associated with events by detecting the dynamics of the tweets. Association Rule Mining (*ARM*) can find the probability of co- existence of tweets' hashtags. Firstly, in [1] *ARM* was used to analyse tweets on the same topic over consecutive time periods t and *t+1*. Rule Matching (*RM*) was later employed to detected changes in patterns such as '*emerging (EM)*', '*unexpected consequent (UnxCs)*' and '*unexpected conditional (UnxCn)*', '*new (N)*' and '*dead (D)*' rules in tweets. This is obtained by setting a user-defined *Rule Matching Threshold (RMT)* to match rules in tweets at time *t* with those in tweets at *t + 1* in order to ascertain rules that fall into the different patterns (as presented in Fig.8). The proposed methodology was coined *TRCM* (*Transaction-based Rule Change Mining)*. All the detected rules in *TRCM* were linked to real life events and news reports.

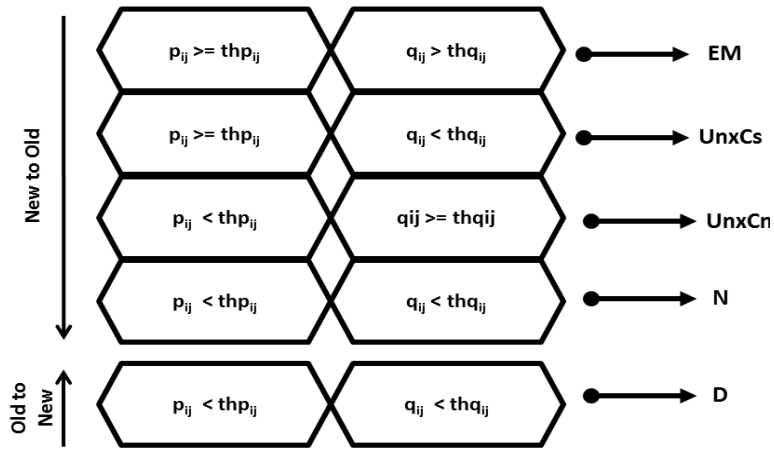

**Fig.8. *TRCM* Rules**

In [37] Subsequently, *TRCM* was utilized to discover the *rule trend* of tweets' hashtags over a consecutive period. *Time Frame Windows (TFWs)* was created (as shown in Fig. 9) to describe different rule evolvement patterns which can be applied to evolvements of news and events in reality. This concept of *TRCM* was named *TRCM-RTI* (*Transaction-based Rule Change Mining-Rule Type Identification*).

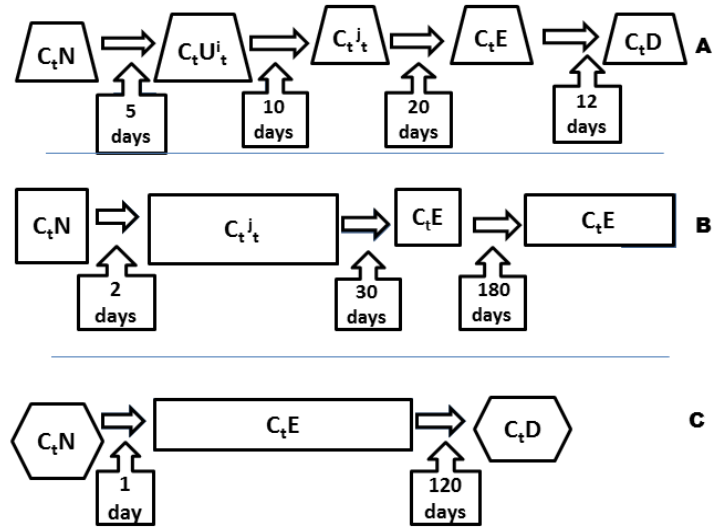

**Fig.9. Different Sequences of Time Frame Windows (*TFWs*)**

Time frame window also serve as a means of calculating the lifespan of specific hashtags on Twitter. Using the experimental study result in [37], it was substantiated that the lifespan of tweets' hashtags can be related to evolvements of news and events in reality. The *TRCM* techniques [1] and [37] can be said to be the first time *ARM* was used to mine Twitter data. This therefore opens up the area for further research as *ARM* can be fine-tuned to be used to mine data on other social networks sites for information retrieval and knowledge acquisition.

Finally different approaches covered in this survey are listed in Table.1.

## 9 Conclusion and Future Work

Different data mining techniques have been used in social network analysis as covered in survey. The techniques range from unsupervised to semi-supervised and supervised learning methods. So far different levels of successes have being achieved either with solitary or combined techniques. The outcome of the experiments conducted on social network analysis is believed to have shed more light on the structure and activities of social network. The diverse experimental results have also confirmed the relevance of data mining techniques in retrieving valuable information and contents from huge data generated on social network. Future survey will tend to investigate novel state-of-the-art data mining techniques for social network analysis. The survey will compare similar data mining tools and recommend the most suitable tool(s) for the dataset to be analysed.

Different data mining techniques covered in this survey are listed in Table.1. The table also contains the approaches employed, the experimental results and the dates and authors of the approaches.

**Table.1 List of Data mining Techniques currently in Used in Social Network Analysis.**

| Approach | Tools | Experiments | Authors/dates |
|---|---|---|---|
| Graph Theoretic | Centrality measure | Inspects representation of power and influence that forms clusters and cohesiveness. | Burts (2005) Borgatti & Everett (2006) |
| | Parameterized centrality metric | Studies the network structure and to rank nodes connectivity. | Ghosh & Leman (2011) |
| | a-centrality | Measures the number of alleviated paths that exist among nodes. | Bonacich & Lloyd (2001) |
| Community Detection (hierarchical clustering) | Vertex clustering | Measures pairwise length between vertices. | Papadopoulos et al (2012) |
| | Structural equivalence measures | Detects friendship structure on social network based on shared behaviour. | Fletcher et al (2011) |
| | Random-walk-based similarity -walktrap | Detects a hierarchical structure of small communities | Pons and Latapy, (2005) |
| Recommender System | *CF* (Collaborative filtering) | Exploits association among users by way of item recommendation. | Liu & HJ Lee (2010) |
| | Content-based | Explores the structures of the data to produce recommendations. | Adomavicius & Tuzhilin, (2005) |
| | Hybrid approach - EntreeC | Suggest recommendations by combining CF and content-based recommendations. | Burke (2002) |

|  | Hierarchical agglomerative clustering | Use to suggest forthcoming conferences or journals in which researchers. | Pham et al (2011) |
|---|---|---|---|
| Semantic Web | Friend of a Friend (FOAF) | Used to explore how local and global community level groups develop and evolve in large-scale social networks on the Semantic Web. | Zhou et al (2011) |
|  | Semantic Web-based Social Network Analysis Model | Combined with conventional outline of the semantic web to create the ontological field library of social network analysis in order to attain intelligent retrieval of the Web services. | Ruan et al (2014) |
|  | VoyeurServer | Improved on the open-source Web-Harvest framework for the collection of online social network data for studying structures of trust enhancement and online scientific association. | Murthy et al (2013) |
| Opinion Analysis | Aspect-Based/Feature-Based | To identify positive or negative opinion sentences in product reviews. | Hu & Liu (2004) |
| Opinion Formation | Homophily Clustering | Used to link same opinion under the same nodes. | Lynn Smith-Lovin & Cook, (2001) Jackson M, 2010 |
| Opinion Definition and Opinion Summarization | Support Vector Machine (SVM/linear kernel) | Used to learn the polarity of neutral examples in documents. | Ku et al (2006) |
| Sentiment Orientation (SO) | hierarchical classification technique | Used to improve the performance of mood classification. | Keshtkar, & Inkpen (2009) |
| Product Ratings and Reviews | Matrix factorisation method | Used to increase rating predictions and estimate accurate strengths of trust associations | Au Yeung and Iwata (2011) |

|  |  | within the same period. |  |
|---|---|---|---|
|  | Social-Union method | Used to exploit multi-modal social networks that provide item recommendations in social rating networks (SRNs). | Symeonidis et al (2011) |
|  | *RnR* Reviews and Ratings | It employs user-input- oriented-system to develop relative new reviews. | Rahayu, 2010 |
| Aspect Rating Analysis | Latent Aspect Rating Analysis (LARA) | Used to determine every reviewer's latent score on each aspects and the relevant influence on users when making final decision. | Wang et al (2010) |
| Unsupervised Classification | part of speech tagging | Used to rate a review as 'thumbs up' or 'thumbs down'. | Santorini (1995) |
|  | PMI-IR/ average semantic orientation of phrase. | Used to approximate semantic orientation of phrases. | Turney (2001) |
|  | EM-based & constrained-LDA | Used to cluster aspect phrases into aspect categories. | Zhai et al, (2011) |
|  | Bootstrapping | It utilizes obtainable primary classifier to make labelled data which a supervised process can build upon. | Kaji & Kitsuregawa (2006), Riloff et al (2003), Wiebe & Riloff (2005) |
| Semi-supervised Classification | Semi-supervised lexical classification | Used to integrate lexical knowledge into supervised learning. | Sindhwani & Melville (2008) |
|  | Polarity detection | Use to solve label propagation in graphs. | Rao &Ravichandran (2009) |
|  | Metric labelling/SVM regression | Used to compare similarity measure. | Pang & Lee (2005) |
| Supervised Classification | Multiple facts base | Used to create graph with nodes and links which represents adjectives and similarity. | Turney (2002) |
| Topic Detection and Tracking (TDT) | *SVM* (Support vector machine) | Used to train Twitter hashtags metadata to predict the political alignment of twitter users. | Conover et al, (2011) |
| | incremental online clustering/ Naïve Bayes-Text classifier | Used to distinguish between fastest growing real world events contents and non-events contents on Twitter. | Becker et al (2011) |
| | browser plug-in script & customizable web interface | Used to identify relevant Twitter content for planned events. | Becker et al (2011b) |
| | *LDA* (Latent Dirichlet Allocation), *Doc-p* (Document-Pivot Topic Detection, *GFeat-p* (Graph-based Feature-Pivot Topic Detection), *FPM* (Frequent Pattern Mining), *SFPM* (Soft Frequent Pattern Mining) and *BNgram* | Used to detect real world events on Twitter. | Aiello et al (2013) |
| | *Hotstream* | Used for detecting and tracking breaking news in Twitter. | Phuvipadawat & Murata (2010) |
| | *FSD* (First Story Detection) | Used to detect predictable and unpredictable events using real-time indication from Wikipedia and Twitter data streams. | Osborne et al (2012) |
| | *EDCoW* (Event Detection with Clustering of Wavelet-based Signals) | Used to cluster words to form events with a modularity-based graph partitioning method. | Weng and Lee (2011) |

| | Latent Dirichlet Allocation topic inference model | Used to ascertain sudden increases on the mention of specific hashtags. | Cordeiro (2012) |
|---|---|---|---|
| | *TRCM* (Transaction-based Rule Change Mining). | Used to link the evolvement of Association Rules present in tweets' hashtags with evolvements of real life news/ events. | Adedoyin-Olowe et al (2013) |